\documentclass[a4paper,twocolumn]{article}
\usepackage{latexsym}
\usepackage{graphics}
\usepackage{epsfig}
\usepackage{multirow}
\usepackage{amssymb}
\usepackage{times}

\newcommand{\etmiss}{\mbox{\ensuremath{\not \hspace{-0.1cm}E_T}}}

\DeclareOldFontCommand{\tt}{\normalfont\sf\small}{\mathtt}

\newcommand{\authorrunning}[1]{\edef\authorrun{#1}}
\newcommand{\titlerunning}[1]{\edef\titlerun{#1}}
\usepackage{fancyhdr}
\begin{document}

%


\title{Prospects for Observing an Invisibly Decaying Higgs Boson in the $\mathbf{t \bar t
H}$ Production at the LHC}

\titlerunning{Prospects for Observing an Invisibly Decaying Higgs Boson\ldots}

\authorrunning{{B. P. Kersevan, M. Malawski, E. Richter-Was}}

\author{
Borut Paul Kersevan \\
\small Faculty of Mathematics and Physics, University of Ljubljana,
 Jadranska 19, SI-1000 Ljubljana, Slovenia. \\ \small and  \\
\small  Jozef Stefan Institute, Jamova 39, SI-1000 Ljubljana, Slovenia.\\[10pt]
Maciej Malawski\\
\small Institute of Computer Science, University of Mining and Metallurgy,\\
\small 30-059 Krakow, al. A. Mickiewicza 30, Poland.\\[10pt]
El\. zbieta Richter-W\c{a}s\thanks{Supported in part by  Polish Government grant KBN 2P03B11819.}\\
\small Institute of Computer Science, Jagellonian University, 30-072 Krakow,
 ul. Nawojki 11, Poland.\\ \small and\\
\small Institute of Nuclear Physics, 31-342 Krakow, ul. Radzikowskiego 152, Poland.
}





\maketitle

\abstract{\it The prospects for observing an invisibly decaying Higgs boson in the $t \bar tH$
production at LHC are discussed. An isolated lepton, reconstructed hadronic
top-quark decay, two identified b-jets and large missing transverse energy are
proposed as the final state signature for event selection. Only the Standard
Model backgrounds are taken into account.  It is shown that the $t\bar t Z$, $t
\bar t W$, $b \bar b Z$ and $b \bar b W$ backgrounds can individually be
suppressed below the signal expectation.  The dominant source of background
remains the $t \bar t$ production. The key for observability will be an
experimental selection which allows further suppression of the contributions
from the $t \bar t$ events with one of the top-quarks decaying into a tau
lepton. Depending on the details of the final analysis, an excess of the signal
events above the Standard Model background of about 10\% to 100\% can be
achieved in the mass range $m_H= 100-200$~GeV.\bigskip \hrule}



\section{Introduction}

While several production and decay modes of the Higgs boson have already been
studied in the past \cite{ATL-PHYS-TDR}, the invisibly decaying Higgs boson has
not yet been exhaustively discussed in the searches scenarios of the LHC
experiments.  There are however many different and reasonable theoretical ideas
which implicate an invisibly decaying Higgs boson. These motivations include
models with light neutralinos, spontaneously broken lepton number, radiatively
generated neutrino masses, additional single scalar(s), or right handed
neutrinos in the extra dimensions.  For a nice recent overview see
e.g. \cite{SPMartin}.

Some theoretical studies have already some time ago addressed the question of
how to look for the evidence of an invisible Higgs decay.  In \cite{DPRoy94} the
use of the $WH/ZH$ production mode was suggested and roughly evaluated. 
This analysis have been recently revised in \cite{Godbole03}. The
observation of the invisibly decaying Higgs boson in the associated $t\bar t H$
production has been proposed in \cite{Gunion93}. Prospects for the observability
of this decay mode in the Vector Boson Fusion production has been proposed and
evaluated in \cite{Dieter00}.  The aforementioned options have been recently
revisited in \cite{LesHouches01}, where the results from the more
experiment--specific analyses were reported.

In this paper the prospects for observation of an invisibly decaying Higgs in
the $t \bar t H$ production are revised. The evidence will be an excess of very
exclusively selected events with a single isolated lepton, large missing
transverse energy, two identified b-jets and one reconstructed top-quark in the
hadronic decay mode.  Such signature requires very dedicated work on
understanding the systematic sources originating in both physics and detector
simulation.  Currently, all these aspects can certainly not be covered. The aim
of this paper is rather to evaluate possible sources of the Standard Model
backgrounds and to identify the dominant contributions thereof.

\section{Signal and Background Processes}

The proton-proton collisions at 14 TeV centre-of-mass energy are simulated,
using the matrix element based generator {\tt AcerMC} \cite{AcerMC} and general
purpose generators {\tt PYTHIA} \cite{Pythia6.2} and {\tt HERWIG}
\cite{Herwig6.3} for event generation.

The signal events, $gg, q \bar q \to t \bar t H $, are generated with the {\tt
PYTHIA} event generator. No model for the invisibly decaying Higgs is assumed;
the only postulate is that the Higgs boson is {\it invisible} to the
detector. The latter is equivalent to assuming a 100\% branching ratio of the
Higgs boson to invisible particles and its coupling to the $t \bar t$ pair is set
equal the Standard Model prediction. In addition, any assumptions on the
mass/spin of the invisible decay products are omitted. The signal production in
the mass range 100-200 GeV is analysed.
 
For simulation of the computationally demanding $2 \to 4$ background
processes the presented study benefits from the availability of the
matrix element implementations and efficient phase-space modeling in
the {\tt AcerMC} generator. Events generated with the matrix elements
of {\tt AcerMC} are further evolved through the QCD shower algorithms
and eventually hadronised using the shower evolution provided by {\tt
PYTHIA}. In addition, the top-quark decays in the matrix element
processes generated by {\tt AcerMC} are handled by {\tt PYTHIA}.  The
{\tt CTEQ5L} parton density functions \cite{cteq5l} and the default
settings of the initialisation parameters for {\tt PYTHIA} and {\tt
HERWIG} are used. The cross-sections for signal and background
processes are specified in Table~\ref{T2:a}.

The proposed analysis relies on identifying the top-quark pair production in the
association with the invisibly decaying object and the lepton-hadron\footnote{
{\it lepton-hadron} denotes one top quark decaying $t
\to Wb \to q \bar q b$ and the other $t \to Wb \to \ell \nu b$, where
$\ell$ stands for electron or muon.} decay mode of the top-quark pair, where an
isolated lepton will trigger an experiment. In the initial step of the events
selection one requires two identified (tagged) b-jets, at least two additional
jets in the central detector region and a large missing transverse energy. The
possible background processes are those which involve a top-quark pair or
b-quark pair production associated with the W or Z-boson.

$\mathbf{gg, q \bar q \to t \bar t}$ : This irreducible continuum background is
generated both with {\tt PYTHIA} and {\tt HERWIG} generators. As the
implementations of the QCD showering/hadronisation models are different in {\tt
PYTHIA} and {\tt HERWIG}, it is considered as very interesting to study how the
consistency of the estimates from both event samples, while the same initial
cross-section is assumed\footnote{The {\tt PYTHIA} cross-section value is used
throughout the analysis. The {\tt HERWIG} cross-section prediction is lower by
$\sim$17\%, mainly due to the different implementation of $\alpha{\rm QCD}$.}.

\begin{table}
\small 
\newcommand{\lstrut}{{$\strut\atop\strut$}}
  \caption {\small Cross-section for signal and background processes. Branching
  ratios are included only for hard-processes $W \to \ell \nu$, $Z \to \nu \nu$
  and $Z/\gamma^* \to \ell \ell$ decays (3 families of neutrinos and two
  families of leptons). For W-bosons originating in top-quark decays all decay
  channels are allowed. For the $Z/\gamma^* \to \ell \ell$ the cutoff $m_{\ell
  \ell} >$ 10 GeV is set. In the signal simulation the strength of the Higgs
  coupling to the top-quark is assumed to be the equal to the Standard Model
  prediction.
\label{T2:a}} 
\vspace{2mm}  
\begin{center}
\begin{tabular}{lll}
\hline\noalign{\smallskip}
Process & Generator & $\sigma$ ($\sigma \times BR$)  \\
\noalign{\smallskip}\hline\noalign{\smallskip}
$t \bar t H$    & {\tt PYTHIA} &  \\
$m_H = 100$ GeV &   &   910 fb    \\
$m_H = 120$ GeV &   &   520 fb    \\
$m_H = 140$ GeV &   &   320 fb    \\
$m_H = 160$ GeV &   &   210 fb    \\
$m_H = 200$ GeV &   &   100 fb    \\
\noalign{\smallskip}\hline\noalign{\smallskip}
$ t \bar t Z$, $Z \to \nu \nu$ &  {\tt AcerMC} &  190 fb\\
\noalign{\smallskip}\hline\noalign{\smallskip}
$t \bar t $&    {\tt PYTHIA}, {\tt HERWIG}   & 490 000 fb  \\
\noalign{\smallskip}\hline\noalign{\smallskip}
$t \bar t W$, $W \to \ell \nu$  &  {\tt AcerMC}  &    140 fb ($\times$ 3) \\
\noalign{\smallskip}\hline\noalign{\smallskip}
$b \bar b W$, $W \to \ell \nu$  &  {\tt AcerMC}  & 73 000 fb   \\
\noalign{\smallskip}\hline\noalign{\smallskip}
$ b \bar b Z$, $Z/\gamma^* \to \ell \ell$ &  {\tt AcerMC} & 61 400 fb    \\
\noalign{\smallskip}\hline
\end{tabular}
\end{center}
\end{table}

$\mathbf{gg, q \bar q \to t \bar t Z, Z \to \nu \nu}$ : 
This irreducible resonant background is generated with the {\tt AcerMC} matrix 
element generator.

$\mathbf{q \bar q \to t \bar t W, W \to \ell \nu}$ :
 This reducible background
is generated with the {\tt AcerMC} matrix element generator. In the implemented
matrix element the W-boson from the hard process is forced to decay into a
lepton and neutrino. As the required lepton can also be produced in the
semi-leptonic top-quark decays, the hard-process W-boson could thus also decay
hadronically; consequently the generated background is multiplied by an
combinatorial factor three in the final results.  This approximation is
acceptable under the assumption that the acceptance is roughly comparable for
events involving either a leptonic decay of the W-boson from the hard process or
a leptonic decay of a W-boson from top-quark decays.  This is an acceptable
assumption as already the initial cross-section for this process is comparable
with the signal values and the irreducible $t \bar t Z$ background
predictions. Even with the combinatorial factor three included, the $ t \bar t W$
background contribution is expected to be on the order of the 
$t \bar t Z$ contribution at most.

$\mathbf{gg, q \bar q \to  b \bar b Z/\gamma^*, Z/\gamma^* \to \ell \ell \oplus {\mathbf \rm  jets}}$ : 
This reducible
background is generated with the {\tt AcerMC} matrix element generator. The $Z/\gamma^*$
is required to decay into a lepton pair.  This background process is
considered to reproduce quite reliably the estimates from the inclusive 'Z-boson 
$\oplus$ jets' production. Recent study in \cite{hep-ph/0203148} has shown that the rates for
$Z b \bar b$ events agree within 10\% with the predictions of the more inclusive
approach of generating $q \bar q \to Z$ hard processes and invoking parton
shower afterward. Requiring a reconstruction of one top-quark in the hadronic
mode, large missing transverse energy and vetoing additional lepton will
strongly suppress this background.

$\mathbf{q \bar q \to b \bar b W, W \to \ell \nu \oplus {\mathbf \rm 
jets}}$ : This reducible background is generated with the {\tt
AcerMC} matrix element generator. The W boson is required to decay to
a lepton and neutrino.  This background represents the lowest limit of
what is expected from the 'inclusive W $\oplus$ jets'
production. Recent study in \cite{hep-ph/0203148} has shown that the
more inclusive, parton shower based estimates for events with two
b-jets and one isolated lepton, are not exceeding the matrix element
results by more than a factor two. Requiring reconstruction of one
top-quark in the hadronic mode and a large transverse missing energy
will suppress it strongly.

The list presented above, although quite exhaustive already, does not include
different reducible backgrounds with one or two misidentified b-jets.  From the
experience of several studies done in \cite{ATL-PHYS-TDR}, one does not expect
these backgrounds to contribute more than 20-30\% of the respective backgrounds
with two true b-quarks.

In the presented analysis about $10^8$ unweighted events were generated for the
$t \bar t$ process with {\tt PYTHIA} and {\tt HERWIG} and about $10^6$ for each 
of the background processes generated with {\tt AcerMC}.

\section{Simplified Detector Simulation}

For the needs of this analysis a simplified version, \cite{AcerDET}, of the fast
simulation/recon\-struction of the ATLAS detector at LHC was used.  It reads
generated events and provides reconstructed experimental observables: isolated
leptons, jets, identified b-jets, transverse energy. Isolated leptons are
reconstructed within the pseudo-rapidity range of $|\eta|<2.5$; the same
pseudo-rapidity coverage is possible for b-jet identification. Jets are
reconstructed within $|\eta|<5.0$. The transverse momentum thresholds are set
for trigger muon to 20~GeV, for trigger electron to 25~GeV and the threshold for
jet reconstruction is set to 15~GeV.  Additional leptons are vetoed with the
threshold of 6~GeV for muon and 10~GeV for electron.

The applied estimates are a 90\% efficiency for lepton
identification and reconstruction, about 80\% efficiency for jet reconstruction
and 60\% efficiency for b-jet identification (with misidentification probability
of 1\% for light jets and 10\% for c-jets).  Resolution of the reconstructed
missing transverse energy components is on the order of 6 GeV.

These performance figures are representative for the low luminosity operation of
the ATLAS detector.  More details about the detector performance and fast
simulation/reconstruction can also be found in
\cite{ATL-PHYS-TDR,ATL-PHYS-98-131}.

\section{Analysis}

The invisibly decaying Higgs boson production in association with two top-quarks
leads to a very distinct signature, namely the large missing transverse energy
and an accompanying top-quark pair.  Requiring a fully or partially
reconstructed top-quark pair will allow for strong suppression of backgrounds
from W or Z production, leaving the $t \bar t$ background as the dominant one.
The (reducible) top-quark background production rate is enormous, the initial
cross-section is by factor $5\times 10^2 - 5 \times 10^3 $ higher then the
signal one.  The only notable distinction between the signal events and the $t \bar t$
events should be a much larger missing transverse energy. Therefore, a
selection which implies accepting the purest possible sample with fully
reconstructed hadronic top-quark decay and partially reconstructed semi-leptonic
top decay is proposed.

The focal point  of the proposed selection is to suppress as much as possible
the contribution from events with one $t \to \ell \nu b$ and another 
 $t \to \tau \nu b$ decay, which results in presence of lepton and a
a large transverse missing energy in the event.

\begin{figure}[ht]
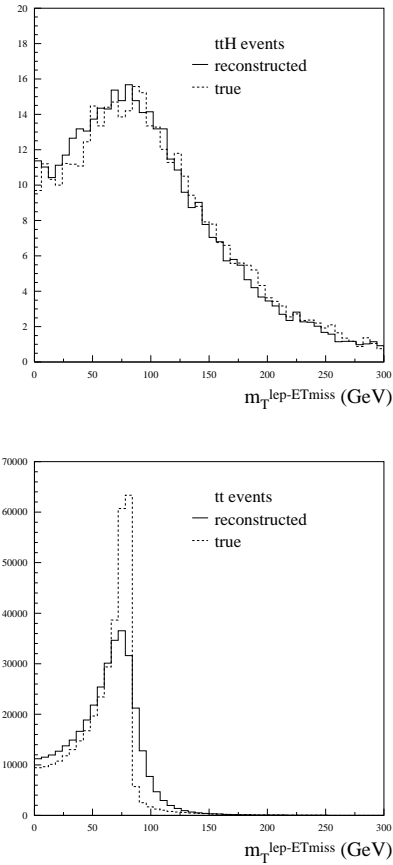

\begin{center}
     \epsfig{file=fig.1a,width=6.0cm}
     \epsfig{file=fig.1b,width=6.0cm}
\end{center}
\caption{
\small
Reconstructed transverse mass of the lepton and \etmiss~ system in the $t
\bar tH$ events (top plot) and in the $t \bar t$ events (bottom plot).  The
dashed line denotes the distributions calculated from the true invisible energy
of the primary products of W boson decays in these events, obtained by using the
generator level information.  The distributions are normalised to the number of
events expected for an integrated luminosity of $30 fb^{-1}$.
\label{FS3:a}} 
\end{figure}

\begin{itemize}
\item
An isolated lepton from the semi-leptonic decay of the one top-quark is required
to provide trigger for these events. In addition, a veto on events
with an additional isolated lepton is set in order to 
suppress the $Z/\gamma^* b \bar b $ background.
\item
Both b-jets have to be identified (tagged), which efficiently reduces background
from the inclusive Z- or W-boson production.
\item
One top-quark reconstructed in the hadronic decay mode $t \to jjb$ is
required. The best $jjb$ combination is chosen from the set of possible
permutations, the criteria being $m_{jj}=m_W \pm 15$ GeV and
$|\eta^{jet}|<2.0$. Taking only the central jets for $W \to jj$ reconstruction
reduces the fraction of events with ``fake'' reconstruction of the `` $W \to
jj$'', where the jets originate in the initial or final state QCD radiation and
not in the W-boson decay. We apply a W-mass constraint in order to re-calibrate
the four-momenta of jets as this optimises the resolution of the reconstructed
$jjb$ system. The $jjb$ system is considered to reconstruct a top-quark if
$m_{jjb}=m_t \pm 25$~GeV.
\item
It is not feasible, without further assumptions, to require a full
reconstruction of the semi-leptonically decaying top-quark in signal events. For
such a reconstruction one needs information on the missing transverse energy
from the W-boson decay.  The latter is however not available, as both the
leptonic W-boson decay and the Higgs boson decay itself contribute to the
missing transverse energy in those events.  Instead, we decided to explore
the fact that the expected transverse mass of the lepton and \etmiss~
system, $m_T(\ell,\etmiss)$, is much higher in the signal than in the $t
\bar t$ background events, see Figure~\ref{FS3:a}.
For the $t \bar t$ events, with the missing transverse energy coming
predominantly from the $W \to \ell \nu$ decay, one can observe characteristic
sharp end-point in the \etmiss~ distribution at about the W-boson mass.  The
tail in this distribution is contributed mostly by events with one $W \to \tau
\nu$ decay or with both W-bosons decaying $W \to \ell \nu$.  For selection we
require $m_T(\ell,\etmiss) > 120$~GeV.
\item
A relatively large missing transverse energy of the system, $\etmiss > 150$~GeV, is
required.
\item
The signal-to-background ratio is enhanced by the additional
requirement of the large transverse momenta in the reconstructed system, $ \sum
p_T^{\mathrm{rec}} > 250$~GeV. The $ \sum p_T^{\mathrm{rec}}= \sum p_T^j + p_T^l$ where the sum
runs over the transverse momenta of reconstructed objects from top-quarks
decays: two b-jets, two light jets used for the reconstruction of the $W \to q
\bar q$ decay and an isolated lepton. This further suppresses the backgrounds
where true top quarks are not present, like $b \bar b Z$ and $b \bar b W$.
\item
Finally, further enhancement of the  signal-to-background ratio
can be achieved by the additional
requirement on the cone separation, $R_{jj}$, between jets which were used for the 
$W \to jj$ reconstruction, the $R_{jj} < 2.2$.
\end{itemize}

\begin{table} 
\small 
\newcommand{\lstrut}{{$\strut\atop\strut$}}
  \caption {\small The cumulative acceptances for the specified selection
  criteria. Efficiencies for b-tagging and lepton identification are
  included. The generation of the event samples was discussed in Sect.2. A
  Higgs boson mass of 120 GeV is assumed for signal events. Only the dominant
  background sources are listed.
\label{T3:a}}  
\vspace{2mm}   
\begin{center}
\begin{tabular}{lllll}
\hline\noalign{\smallskip}
Process & $t \bar t H$ & $t \bar t Z$ & $t \bar t$  & $t \bar t$  \\
        &\hspace{-0.3cm} {\tt PYTHIA} &\hspace{-0.3cm} {\tt AcerMC}&\hspace{-0.3cm}  {\tt PYTHIA} &\hspace{-0.3cm} {\tt HERWIG} \\
\noalign{\smallskip}\hline\noalign{\smallskip}
    Trigger lepton &\hspace{-0.3cm} 22\%  &\hspace{-0.3cm} 22\%  &\hspace{-0.3cm} 22\%  &\hspace{-0.3cm} 22\%     \\
\noalign{\smallskip}\hline\noalign{\smallskip}
2 b-jets + 2 jets &\hspace{-0.3cm} 5.0\%  &\hspace{-0.3cm} 4.8\% &\hspace{-0.3cm} 4.9\% &\hspace{-0.3cm} 5.2\%    \\
\noalign{\smallskip}\hline\noalign{\smallskip}
rec. t-quark (jjb) &\hspace{-0.3cm}  2.6\%  &\hspace{-0.3cm} 2.4\% &\hspace{-0.3cm} 2.4\% &\hspace{-0.3cm} 2.6\%     \\
\noalign{\smallskip}\hline\noalign{\smallskip}
$m_T^{\ell,\etmiss}>120$ GeV &\hspace{-0.3cm} 0.87\%  &\hspace{-0.3cm} 0.93\% &\hspace{-0.3cm} $4.1 \cdot 10^{-4}$ &\hspace{-0.3cm}  $5.2 \cdot 10^{-4}$  \\
\noalign{\smallskip}\hline\noalign{\smallskip}
$\etmiss > 150$ GeV &\hspace{-0.3cm} 0.41\%  &\hspace{-0.3cm} 0.53\% &\hspace{-0.3cm} $ 2.3 \cdot 10^{-5}$ &\hspace{-0.3cm} $ 3.7 \cdot 10^{-5}$  \\
\noalign{\smallskip}\hline\noalign{\smallskip}
$\sum p_T^{\mathrm{rec}}>250$ GeV &\hspace{-0.3cm} 0.40\%  &\hspace{-0.3cm} 0.51\% &\hspace{-0.3cm}  $ 2.0 \cdot 10^{-5}$ &\hspace{-0.3cm} $ 3.2 \cdot 10^{-5}$  \\
\noalign{\smallskip}\hline\noalign{\smallskip}
$R_{\mathrm{jj}}<2.2$ &\hspace{-0.3cm} 0.28\%  &\hspace{-0.3cm} 0.35\% &\hspace{-0.3cm}  $ 7.5 \cdot 10^{-6}$ &\hspace{-0.3cm} $ 1.2 \cdot 10^{-5}$  \\
\noalign{\smallskip}\hline
\end{tabular}
\end{center}
\end{table} 

In Table~\ref{T3:a} the selection criteria and cumulative acceptances from
signal and dominant background processes are specified.  The selection cutoff
$\etmiss > 150$~GeV is quite loose and certainly can be optimised
further. The cumulative acceptance for signal events after these cuts is about 0.3\%. The
acceptance is indeed very similar for the $t \bar t H$ signal events at
$m_H=120$~GeV and for the $t \bar t Z$ background events. The cumulative
acceptance for the $t \bar t$ process is of $ 7.6 \cdot 10^{-6}$ for {\tt
PYTHIA} events and $ 1.2 \cdot 10^{-5}$ for {\tt HERWIG} events.

\begin{figure}[tbh]
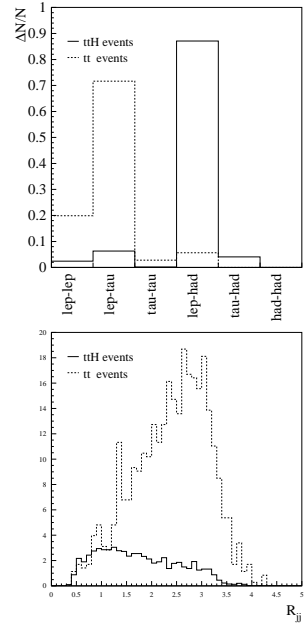

\begin{center}
     \epsfig{file=fig.2a,width=4.3cm}\\
     \epsfig{file=fig.2b,width=4.3cm}
\end{center}
\vspace{-5mm}
\caption{\small
The relative fractions of the $t \bar t$ decay modes are listed for signal and $t \bar t$ 
background simulated  with {\tt PYTHIA} (top plot). 
The $R_{\mathrm{jj}}$ cone separation between jets used in the $W \to jj$ reconstruction;
 the distributions are normalised to the number of events expected for
 an integrated luminosity of $30 fb^{-1}$ (bottom plot).
\label{FS4:a}} 
\vspace{-3mm}
\end{figure}

After performing the selection, about 70\% of the $t \bar t$ events comes from the lepton-tau 
\footnote{ The {\it lepton-tau} label denotes one top quark decaying $t
\to Wb \to \ell \nu b$ and another $t \to Wb \to \tau \nu b$, where
$\ell$ stands for electron or muon. The  {\it lepton-lepton} decay labels events
 with both top quarks decaying  $t\to Wb \to \ell \nu b$.  Finally,
 the {\it lepton-hadron} decay labels events with  one top quark decaying $t
\to Wb \to \ell \nu b$ and another $t \to Wb \to q \bar q  b$ }
decay and 20\% from the lepton-lepton decay of the top-quark pair in the {\tt PYTHIA}
sample with compatible fractions also found in {\tt HERWIG} events.
In these two cases the $jjb$ combination is thus made from the ISR/FSR
jets and not from the true $W \to q \bar q$ decays.  These events could
hopefully be suppressed further by implementing a tau-jet veto and with more
stringent requirements in the $t \to jjb$ reconstruction. The cumulative
acceptance for the $t \bar t$ background is found to be about 50\% higher for
events generated with {\tt HERWIG} than with {\tt PYTHIA} generator. The signal events in 
contrast contain only a $\sim 10\%$ fraction of lepton-tau and lepton-lepton decays; the 
relative fractions of signal and $t \bar t$ background events are shown in Figure~\ref{FS4:a}.

Considering the relative fractions of the tau-lepton events in the signal and background, 
one can assume that the inter-jet cone separation, $R_{\mathrm{jj}}$, might provide
some additional separation power; the $R_{\mathrm{jj}}$ for signal and $t \bar t$ background 
events are given in Figure~\ref{FS4:a}. Subsequently, a loose cut of $R_{\mathrm{jj}}<2.2$ was 
applied; the final efficiencies are listed in Table~\ref{T3:a}.

It can reasonably be assumed that the $R_{\mathrm{jj}}$ cut is the most
sensitive one to the modeling of ISR and FSR jets, which in certain regions of
phase space might not be adequately described by the parton shower generators
such as {\tt PYTHIA} or {\tt HERWIG}. Nevertheless, since the efficiency for the
$R_{\mathrm{jj}}$ cut is nearly identical for the $t \bar t$ backgrounds
produced by {\tt PYTHIA} and {\tt HERWIG}(c.f. Table~\ref{T3:a}), it is assumed
that the cut is robust enough to be included. Due to its clear physics content
it should remain valid also when more accurate simulations of the $t \bar t
\oplus jets$ background become available.

\begin{figure}[hbt]
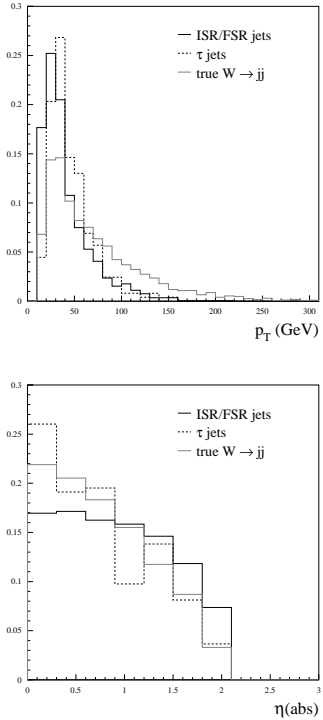

\begin{center}
     \epsfig{file=fig.3a,width=5.cm}\\
     \epsfig{file=fig.3b,width=5.cm}
\end{center}
\caption{\small
Reconstructed transverse momenta, $p_T$ (top plot) and the
rapidity (absolute value),$\eta$ (bottom plot), of the two
light jets used in the $W \to q \bar q$ reconstruction, originating either 
in theISR/FSR, $W \to \tau \nu$ or true $W \to jj$ decays in the $t \bar t$
background. The distributions are normalised to one.
\label{FS5:a}} 
\end{figure}

The distributions of the rapidity and transverse momentum of the two light jets used
in the $W \to q \bar q$  reconstruction in 
contrast do not exhibit a significant difference when originating either in true
$W \to jj$ events, $W \to \tau \nu$ or in the initial or final state radiation but 
for an expected tendency to higher rapidities and lower transverse momenta, typical for the 
ISR/FSR jets. The obtained distributions for the $t \bar t$ background are
shown in Figure~\ref{FS5:a}.

\begin{table} 
\small
\newcommand{\lstrut}{{$\strut\atop\strut$}}
  \caption {\small Expected numbers of events for an integrated luminosity
  of $30 fb^{-1}$ and selection as specified in Table~\ref{T2:a}. 
  Efficiencies for b-tagging and lepton identification
  are included. The {\tt (PY)} and {\tt (HW)} denote the results for the
  $t \bar t$ events generated with {\tt PYTHIA} and {\tt HERWIG}
  respectively. Also shown is the separate contribution to the $t \bar
  t$ background from the lepton-hadron events.
\label{T3:b}}  
\vspace{2mm}  
\begin{center}
\begin{tabular}{lc}
\hline\noalign{\smallskip}
Process & No. of events  \\
\noalign{\smallskip}\hline\noalign{\smallskip}
$t \bar t H$,               &         \\
              $m_H=100$ GeV &  60     \\ 
              $m_H=120$ GeV &  45     \\ 
              $m_H=140$ GeV &  30     \\ 
              $m_H=160$ GeV &  25     \\ 
              $m_H=200$ GeV &  15     \\ 
\noalign{\smallskip}\hline\noalign{\smallskip}
$t \bar t Z$                 &  20      \\ 
\noalign{\smallskip}\hline\noalign{\smallskip}
$t \bar t W  $              &   20  \\ 
\noalign{\smallskip}\hline\noalign{\smallskip}
$t \bar t $ (all)                 &  115 (PY) ,  190 (HW)    \\ 
 (only lepton-hadron)                &   15 (PY) ,   30 (HW)  \\ 
\noalign{\smallskip}\hline\noalign{\smallskip}
$b \bar b W $               &   5 \\ 
\noalign{\smallskip}\hline\noalign{\smallskip}
$b \bar b Z/\gamma^*$       &   5  \\ 
\noalign{\smallskip}\hline
\end{tabular}
\end{center} 
\end{table}

The expected numbers of events for an integrated luminosity of $30 fb^{-1}$ are
given In Table~\ref{T3:b}. Several values of the Higgs boson masses are studied,
while assuming the Standard Model production cross-sections.

Taking as an example the results for the $t \bar t$ background obtained with {\tt
PYTHIA} generator, the signal-to-background ratio is about 39\% for 
Higgs boson mass of 100 GeV and about 9\% for the Higgs mass of 200~GeV.
The total number of expected events from all backgrounds, but the $t
\bar t$ one, is on the level of the signal itself. It is evident that in case 
the ``fake'' reconstructions could be eliminated in the $t \bar t$
events, the signal-to-background ratio could be brought to e.g. 90\% for the
$m_H = 120$ GeV, without changing  thresholds on the \etmiss.
\begin{figure}[ht]
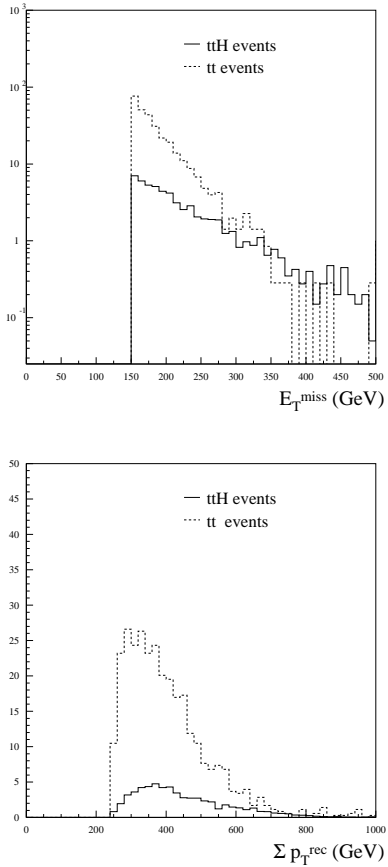

\begin{center}
     \epsfig{file=fig.4a,width=6.0cm}
     \epsfig{file=fig.4b,width=6.0cm}
\end{center}
\caption{\small
Reconstructed missing transverse energy \etmiss~ (top) and sum of the transverse
momenta of reconstructed objects $\sum p_T^{\mathrm{rec}}$ (bottom).
The solid line denotes the $ttH$  signal with $m_H = 120$~GeV, 
the dashed one the $t \bar t$ background prediction.
The distributions before the last selection step specified in Table~\ref{T2:a}
are shown. The distributions are normalised to the number 
of events expected for an integrated luminosity of $30 fb^{-1}$.
\label{FS3:b}} 
\end{figure}

\begin{figure}[ht]
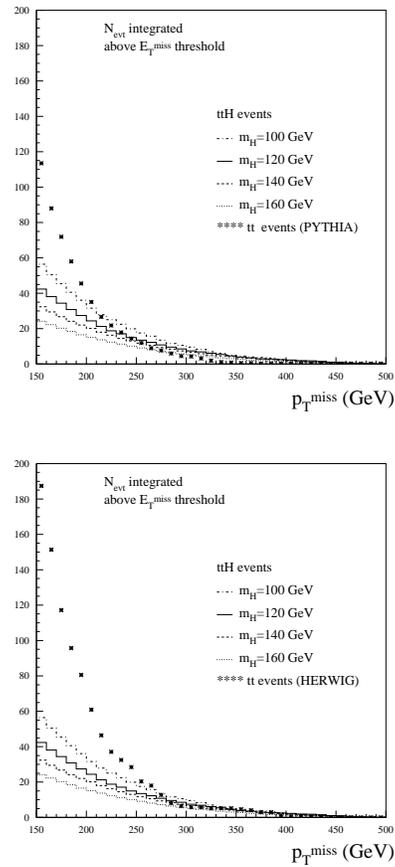

\begin{center}
     \epsfig{file=fig.5a,width=6.0cm}
     \epsfig{file=fig.5b,width=6.0cm}
\end{center}
\caption{\small
Integrated number if events from the $ttH$ signal with $m_H$ = 100, 120, 140,
160~GeV (histograms) and the $t \bar t$ background (stars) as a function of the
\etmiss~ threshold.The results for $t \bar t$ events simulated with {\tt
PYTHIA} (top) and {\tt HERWIG} (bottom) generators are shown. The distributions
are normalised to the number of events expected for an integrated luminosity of
$30 fb^{-1}$.
\label{FS3:c}} 
\end{figure}
  
In Figure~\ref{FS3:b} the expected final (post-selection) distribution of the
\etmiss~ and $\sum p_T^{\mathrm{rec}}$ is shown for the Higgs mass of 120 GeV.  One
can clearly see that the shape of the \etmiss~ distribution is much steeper
for the $t \bar t$ events than for the signal ones. In addition, the shape of
the $\sum p_T^{\mathrm{rec}}$ distribution is different for both classes of events and in
fact the applied threshold 250~GeV is shown to be rather low.  One may also
expect that the shape of the $\sum p_T^{\mathrm{rec}}$ (or analogous) distributions will
be sensitive to the Higgs boson mass, thus allowing for the extraction of that
information with some precision, limited by the signal-to-background ratio.

One can certainly increase the signal-to-background ratio by rising thresholds on
the $R_{\mathrm{jj}}$,\etmiss, $\sum p_T^{\mathrm{rec}}$ or other reconstructed visible observables,
like e.g. the total transverse momenta of the top-quark and lepton system,
$P_T^{\mathrm{rec}}$, or transverse momenta of the top-quark, $p_T^{top}$. Taking as the
reference the distributions shown in Figure~\ref{FS3:c}, increasing the threshold on
the \etmiss~ to about 250~GeV would result in increasing the
signal-to-background ratio (only $t \bar t$ background) to about one, while
still keeping the expected number of signal events to about 20. 

One can also increase signal-to-background ratio by further optimising the 
selection criteria, e.g. by asking for an isolation in the azimuthal angle of the
\etmiss~ direction from the reconstructed jets, reconstructed top-quark or the
isolated lepton.  It has been checked that the moderate isolation requirement,
$\delta \phi > 0.4$, can improve signal-to-background ratio by 20-30\% at the
price of reducing signal rates by a comparable amount.

Nevertheless, in our opinion, the key point for the signal observability remains
the experimental efficiency for reducing the fake $W \to q \bar q$
reconstruction and thus the contribution from the $t \bar t$ events with
lepton-tau and lepton-lepton decays.

\section*{Conclusions}

The prospects for observing the invisibly decaying Higgs boson in the $t \bar
tH$ production at LHC were discussed. The proposed analysis required one
top-quark reconstructed in the hadronic decay mode, an isolated lepton
(electron, muon) from the decay of the second top-quark and a large missing
transverse energy.  Evidence for signal would be an observation of an excess of
such events above the background.  Expected excess can be on the level from 10\%
to even 100\% or more, depending on the required threshold on the missing
transverse energy and on the assumed Higgs boson mass. It can be expected that
some sensitivity to the Higgs boson mass could be revealed by the hardness of
the reconstructed visible part of the event, the $\sum p_T^{\mathrm{rec}}$, $P_T^{\mathrm{rec}}$
or similar distributions.  The signal observability should not degrade
significantly for the high-luminosity operation of the detectors. Thus, the
sensitivity to the signal is expected to rise with the increasing collected
integrated luminosity.

The availability of the matrix element implementations for $t\bar t Z$, $t \bar
t W$, $t\bar t Z$ and $t \bar t W$ processes in the {\tt AcerMC} generator
allowed to conclude that the total contribution from these background processes
could be kept on the level or below the signal one.

The do\-mi\-nant Stand\-ard Model back\-ground comes from the $t \bar t$ production with
one top-quark decaying semi-lep\-ton\-ically into electron or muon and the second
one into tau-lepton. It was also shown, by comparing results for the $t \bar t$
background generated with {\tt PYTHIA} and {\tt HERWIG} Monte Carlo, that for
the final estimate one would have to study very carefully systematics from the
showering, hadronisation and decays models.  The key to reduce the $t \bar t$
background further will be the purest possible reconstruction of the top-quark
hadronic decays ($ t \to q \bar q b$), thus eliminating the events with the
top-quark decaying to tau-lepton ($ t \to \tau \nu b$).

It was concluded that the final optimisation of the observability potential
demands much more sophisticated experimental analysis than foreseen in the scope
of this paper. Rather than increasing threshold on the required missing
transverse energy one should aim for the best possible suppression of the
contribution from the tau-lepton events of the top-quark pair decays.

In Table~\ref{T4:a} a comparison between the sensitivity to the invisible Higgs
production in the $t \bar t H$ channel and in the $q q \to q q H$ vector boson
fusion (VBF) published in \cite{LesHouches01} is given in terms of the
sensitivity of the $\xi^2$ parameter: 
$$\xi^2 = \frac{\sigma{\rm (t \bar t H)}}
{\sigma{\rm (t \bar t H)}_{\rm SM}} \times {\rm Br(H \to inv)}.$$

The parameter $\xi^2$, as defined in \cite{LesHouches01} and presented in
Table~\ref{T4:a}, serves as an estimate of the branching fraction\\ ${\rm Br(H \to
inv)}$. It is derived from the fact that since the true $\sigma{\rm (t \bar t H)}$
is also not known {\it a priori}, one is restricted to measuring the $\sigma{\rm
(t \bar t H \to inv)}= \sigma{\rm (t \bar t H)} \times {\rm Br(H \to inv)}$ and
thus the ${\rm Br(H \to inv)}$ estimate has to be scaled by the ratio between
the unknown cross-section and the Standard Model prediction $\sigma{\rm (t \bar
t H)}_{\rm SM}$. The Table~\ref{T4:a} lists the sensitivity limits of the
$\xi^2$ which can be probed at 95\% confidence level.  In the table the VBF
limits have been re-scaled to the luminosity of $\rm 30 fb^{-1}$; in case the
data point was not provided the nearest value was taken.

The presented limits of $\xi^2$ do not contain the estimates of systematic uncertainties since 
the level of uncertainty about the background predictions is in our opinion  still too
large, as it is evident from the difference between {\tt PYTHIA} and {\tt HERWIG} predictions for
the $t \bar t$ background.  It might well be, that for more firm background estimates one might   
have to wait for the availability of NLO Monte-Carlo generators and/or tuning on the data itself. 
\begin{table} 
\small
\newcommand{\lstrut}{{$\strut\atop\strut$}}
\caption {\small
Expected sensitivities of $\xi^2$ for an integrated luminosity
  of $30 fb^{-1}$ and selection as specified in Table~\ref{T2:a}. 
  In the first column the complete {\tt PYTHIA} background prediction is considered and
  while in the second column only the lepton-hadron $t \bar t$ decays are included. The third
  column lists the values from the VBF analyses given in \cite{LesHouches01}, re-scaled for 
  comparison to the integrated luminosity of $30 fb^{-1}$.  
\label{T4:a}}  
\vspace{2mm}  
\begin{center}
\begin{tabular}{lccc}
\hline\noalign{\smallskip}
Process & $\xi^2$[\%]($t\bar tH$) & $\xi^2$[\%]($t \bar t H$) &$\xi^2$[\%]($VBF$)  \\ 
        & all $t \bar t$ & (lep-had) $t \bar t$ &  \\ 
\noalign{\smallskip}\hline\noalign{\smallskip}
$t \bar t H$, &       & &         \\
$m_H=100$ GeV & 42.2  & 26.5 & 12.1  \\ 
$m_H=120$ GeV & 55.7  & 27.4 & 10.3  \\ 
$m_H=140$ GeV & 75.4  & 47.4 & 9.8  \\ 
$m_H=160$ GeV & 95.6  & 60.2 & 9.9  \\ 
$m_H=200$ GeV & 154.3 & 97.1 & 10.7   \\ 
\noalign{\smallskip}\hline
\end{tabular}
\end{center}
\end{table}
It is nevertheless evident that even if an efficient way to reject events with {\it fake}
$W \to jj$ reconstruction (topological selection,  tau-jet veto) is found 
the potential for invisible Higgs detection in the  $t \bar t H$ channel is
about a factor two to three less weaker than with the VBF channel \cite{LesHouches01} 
in the low-mass Higgs region while in the high-mass region the clear VBF dominance is evident. 
One has to stress, however that the $t \bar t H$ channel does not require the implementation of 
an efficient forward jet trigger essential for the VBF studies as stated in \cite{LesHouches01} 
and that with more stringent cut optimisation it might still  be  possible to
significantly increase the sensitivity listed in Table~\ref{T4:a}. 

The associated Higgs production $t \bar t H$ already turned to be very powerful
for the $H \to b \bar b$ decay mode \cite{ActaB30}. Establishing the
observability in the same production mode and the complementary decay channel
(if $H \to inv$ is open the $H \to b \bar b$ is suppressed) will make this
search very interesting in the range of the intermediate strength of both
decays.




\end{document}